# Artificial Intelligence to Assist in Exclusion of Coronary Atherosclerosis during CCTA Evaluation of Chest-Pain in the Emergency Department: Preparing an Application for Real-World Use

Richard D. White, Barbaros S. Erdal, Mutlu Demirer, Vikash Gupta, Matthew T. Bigelow, Engin Dikici, Sema Candemir, Mauricio S. Galizia, Jessica L. Carpenter, Thomas P. O'Donnell, Abdul H. Halabi, Luciano M. Prevedello

*Abstract*— Coronary Computed Tomography Angiography (CCTA) evaluation of chest-pain patients in an Emergency Department (ED) is considered appropriate. While a "negative" CCTA interpretation supports direct patient discharge from an ED, labor-intensive analyses are required, with accuracy in jeopardy from distractions. We describe the development of an Artificial Intelligence (AI) algorithm and workflow for assisting interpreting physicians in CCTA screening for absence of coronary atherosclerosis.

The two-phase approach consisted of: (1) *Phase 1* - focused on development and preliminary Testing of an algorithm for vessel-centerline extraction classification in a balanced study population (n = 500 with 50% disease prevalence) derived by retrospective random case selection; and (2) *Phase 2* - concerned with simulated-clinical Trialing of developed algorithm on a per-case basis in a more "real-world" study population (n = 100 with 28% disease prevalence) from an ED chest-pain series. This allowed pre-deployment evaluation of the AI-based CCTA screening application which provides vessel-by-vessel graphic display of algorithm inference results integrated into a clinically capable viewer. Algorithm performance evaluation used Area Under the Receiver-Operating-Characteristic Curve (AUC-ROC); confusion matrices reflected ground-truth *vs* AI determinations.

The vessel-based algorithm demonstrated strong performance with AUC-ROC = 0.96. In both *Phase 1* and *Phase 2*, independent of disease prevalence differences, negative predictive values at the case level were very high at 95%. The rate of completion of the algorithm workflow process (96% with inference results in 55-80 seconds) in *Phase 2* depended on adequate image quality.

There is potential for this AI application to assist in CCTA interpretation to help extricate atherosclerosis from chest-pain presentations.

Corresponding author: White.Richard1@mayo.edu

R. D. White, B. S. Erdal, M. Demirer, and V. Gupta are in the Department of Radiology, Mayo Clinic, Jacksonville, FL 32224, USA (email: White.Richard1@mayo.edu; Erdal.Barbaros@mayo.edu; Demirer.Mutlu@mayo.edu; Gupta.Vikash@mayo.edu).

M. T. Bigelow, E. Dikici, S. Candemir, J. L. Carpenter, and L. M. Prevedello are in the Department of Radiology, The Ohio State University College of Medicine, Columbus, OH 43210, USA (email: Matthew.Bigelow@osumc.edu; Engin.Dikici@osumc.edu; Sema.Candemir@osumc.edu; Jessica.Carpenter@osumc.edu; Luciano.Prevedello@osumc.edu).

M. S. Galizia is in the Department of Radiology, University of Iowa, Iowa City, IA 52242, USA (email: Mauricio-Galizia@uiowa.edu).

T. P. O'Donnell is with Siemens Healthineers, Malvern, PA 19355, USA, (e-mail: Tom.Odonnell@siemens-healthineers.com).

A. H. Halabi is with the NVIDIA, Santa Clara, CA 95051, USA, (e-mail: AHalabi@nvidia.com).

## I. BACKGROUND

NEW-ONSET chest pain remains a predominant clinical presentation to an Emergency Department (ED), and it can lead to significant concurrent costs from testing and/or anxiety, as well as future financial and/or time expenses following initially abnormal findings[1,2]; this burden persists despite Acute Coronary Syndrome (ACS) being a relatively infrequent final diagnosis (i.e., in only 17% of ED chest-pain cases[3]). However, due to the clinical or medical-legal consequences of failing to recognize ACS (i.e., in up to 4.4% of ACS cases[3]), ED chest-pain patients continue to be frequently hospitalized even when at little risk (e.g., with low-to-intermediate pretest probability) of "obstructive" Coronary Artery Disease (CAD) (i.e., $\geq$ 50% luminal diameter stenosis)[2,4]. This has fostered opportunity for alternative approaches to chest-pain evaluation, such as Coronary Computed Tomography Angiography (CCTA).

Validated by its high Sensitivity for atherosclerosis detection, as well as its high Negative Predictive Value (NPV) and Specificity supporting disease exclusion, CCTA has become a well-established option for CAD evaluation in symptomatic ED patients[1,2,4,5]. Such use of CCTA in chest-pain patients to effectively eliminate CAD as the cause of symptoms is now endorsed by professional appropriateness guidelines[6], and it is considered especially suitable in the clinical setting of low-to-intermediate CAD probability as is found in most (i.e., over 80%[7]) ED presentations for suspected coronary-related chest pain[6]. Moreover, direct patient discharge from the ED to home based on a "negative" CCTA examination (often defined by absence of obstructive disease[5]) is supported by its: 1. noninvasiveness and patient acceptability; 2. safety and positive outcome projection (e.g., absent cardiac deaths, stable-to-decreased Major Adverse Cardiac Event (MACE) occurrence, and stable-to-decreased repeat visits or admissions following ED discharge); and 3. generic cost-savings (funds and time) to healthcare systems (e.g., shortened and less costly initial visits; reduced post-visit

expenses)[1,2,4,8–11].

However, for the full clinical benefit of this application to be realized, the *complete exclusion of any coronary atherosclerosis,* not only obstructive CAD, within the entire coronary artery tree per chest-pain patient is essential because the presence of even mild degrees of atherosclerotic plaque on CCTA is known to predispose to ACS[12,13]. In addition, if coronary atherosclerosis is detected on CCTA, and quantitation or display of atherosclerotic plaque burden is desired for treatment planning or prognostication[14], accurate delineation of the coincident disease-free arteries, branches, or segments becomes a prerequisite.

Unfortunately, the widespread utilization of this CCTA option in the ED setting is highly restricted by the non-uniform availability of skilled interpreting physicians to support uninterrupted diagnostic activities across multiple sites[10]. In addition, growing pressure from referring clinicians for rapid turn-around of definitive imaging results, in order to expedite ED-patient throughput[15], creates increasing challenges for CCTA interpreters. With rising demands on medical professionals related to more-and-more stressful healthcare delivery environments and communication complexities, frequent disturbances of physician workflow and concentration during a CCTA interpretation are growing prospects for professional mistakes[16,17]. This could potentially result in a falsely negative analysis, and ultimately to an unwanted early and dangerous discharge of a chest-pain patient from an ED due to undetected coronary atherosclerosis[3]. The application of Artificial Intelligence (AI) in medical imaging has the potential to both enhance yield and reduce human error in diagnostic assessments through the assistance it provides to interpreting physicians[18,19].

In recent years, AI-based approaches have been used for the direct detection and characterization of known coronary atherosclerosis on CCTA[20–23]. On the other hand, to our knowledge, there has been no prior application of AI to the direct *exclusion* of atherosclerosis on CCTA for the aforementioned reasons, especially in the ED setting.

In this study, a two-phase approach was used to develop, assess, and prepare an AI algorithm for a "real-world" deployment aimed at prospectively assisting an interpreting physician in accurate and prompt *exclusion of coronary atherosclerosis* on CCTA in ED chest-pain patients. This preparation includes capabilities for concurrent return of inference results to the interpreter graphically within a clinical viewer, on a vessel-by-vessel basis, for adjudication.

## II. Materials and Methods

Following a brief description of the: (1) basics of CCTA imaging methodology used, (2) coronary artery image-processing performed, and (3) AI resources employed, this section describes the two phases leading to the creation, assessment, and pre-deployment of an AI algorithm to assist the interpreting physician in the exclusion of coronary atherosclerosis on CCTA. *Phase 1* was focused on the development and preliminary evaluation of the vessel-based algorithm grounded on a balanced (diseased *vs* non-diseased) study population derived by retrospective random selection of CCTA cases and supported by Data Augmentation (DA). *Phase 2* was concerned with the simulated-clinical Trialing of the developed vessel extraction-based algorithm in a more "real-world" study population comprised of a recent series of consecutive ED cases with chest pain evaluated using CCTA. It allowed a pre-deployment evaluation of the proposed case-based application, which concurrently provides to the interpreter the algorithm inference results by vessel-by-vessel graphic display integrated into a clinically capable viewer.

### A. CCTA Imaging Methodology

All CCTA examinations were performed using American College of Radiology (ACR)-accredited single-source (Definition AS Plus, Edge, or go.Top) or dual-source (Definition Flash or Force) multi-detector systems [Somatom: Siemens Healthineers, Forchheim, Germany], with ECG-based referencing of image-data acquisition[24]. Following standard processes[25], heart-rate control (target $\leq$ 60 beats/minute using intravenous Metoprolol $\leq$ 30 milligram) and vasodilatation (using sublingual Nitroglycerin 0.4 milligram) preceded injector-controlled [Stellant: Medrad, Indianola, PA] administration of a non-ionic/low-osmolar contrast agent (intravenous Iohexol-350 80 ml, followed by saline 50 ml, at 5-6 milliliter/second) [General Electric Healthcare Medical Diagnostics, Chicago, IL]. Emphasizing the "As Low As Reasonably Achievable" principle[26], limitation of radiation exposures was achieved using (in decreasing order of preference): (1) single-beat helical, (2) prospectively triggered sequential, or (3) pulse-modulated retrospectively gated helical acquisitions[24,25].

### B. Coronary Artery Image-Processing

The stepwise CCTA image-processing methodology used per case included: (1) segmentation of the coronary arteries in the entire cardiac CT volume, and (2) use of a Graphical User Interface (GUI)[27,28] for both manual coronary artery analysis and the evaluation of automatic vessel-centerline extractions [CT Cardio-Vascular Engine: Siemens Healthineers, Forchheim, Germany, combined with proprietary technology [29]], which were performed on coronary arteries or branches either with or without atherosclerosis[30] [Appendix A].

### C. AI Resources

AI computational support was provided by multiple servers and workstations with integrated Graphics Processing Units (GPUs) [Quadro GV100, TITAN V (Volta), GTX 1080 Ti, and GTX TITAN X: Nvidia, Santa Clara, CA[31]]. These systems enable technologies (e.g., Clara Medical Imaging[32]), with distributed functionality as a node cluster for processing optimization (e.g., Linux-based Kubernetes v1.8 for managing variable-generation GPUs[33]).

**TABLE I**
*Phase 1* Algorithm Development and Preliminary Evaluation

The vessel-based algorithm development (including Training, Validation and Testing) utilized Mosaic Projection Views (MPVs) of vessel-centerline extractions following the courses of primary coronary arteries or branches which themselves did *vs* did not demonstrate atherosclerosis. If atherosclerosis was present along the vessel extraction, its extent was manually annotated (Plaque-Annotated) as non-obstructive or obstructive plaque[27]. During Training, 6-fold Data Augmentation (DA) of Plaque-Annotated MPVs provided increased representations of atherosclerotic vessels and ultimately improved Training subset balance; a single MPV without DA was used in both Validation and Testing of algorithm performance.

Case-level Testing was conducted following the restoration of each Diseased case by the addition of previously excluded MPVs.

|  |  | Vessel-Based Algorithm Data Subsets | | | |
|---|---|---|---|---|---|
| *Phase 1* Study Population (n = 500) | | *Training* | *Validation* | *Testing* | |
| Expert "Ground Truth" Case Class | Diseased (n = 250 including 125 non-obstructive / 125 obstructive) | 150 | 50 | 50 | - |
| | Normal (n = 250) | 150 | 50 | 50 | - |
|  |  | Algorithm-Development Operations | | | |
| Level of Operation | | *Training* | *Validation* | *Testing* | *Restored-Case Testing* |
| Vessel Extraction (MPV display) | Plaque-Annotated (n = 2,539) | 2,364* | 50 | 125 | - |
| | Atherosclerosis-Free (n = 3,420) | 2,304 | 50 | 1,066 | - |
| Case | Diseased | | | | 50 # & |
| | Normal | | | | 50 |
| Applied to MPVs from Vessel-Centerline Extractions | Resulting from 6-fold DA of 394 Plaque-Annotated MPV per extraction | * | - | - | - |
| | Use of single MPV per extraction | - | X | X | - |
| | Inclusion of extraction MPVs of atherosclerosis-free branches with "upstream" plaque annotations | - | - | - | # (n = 538 added to complete Diseased-case image datasets) |
| | Inclusion of completely atherosclerosis-free extraction MPVs | - | - | - | & (n = 353 added to complete Diseased-case image datasets) |

## *Phase 1: Algorithm Development and Preliminary Evaluation*

### *Study Population:*

With local Institutional Review Board (IRB) approval, a retrospective electronic-record search led to the identification of a standard-of-care experience (5/2013-10/2018) with CCTA for new-onset chest pain in non-hospitalized patients. This experience was reflected in clinical CCTA-examination reports which were generated by either: (1) 1 of 8 interpreting physicians with verification of American College of Cardiology Foundation (ACCF)/American Heart Association (AHA) Cardiovascular Computed Tomography Experience (CCTE) at Level-2[34], but more often by (2) 1 of 2 interpreters with ACCF/AHA CCTE at Level-3[34], as well as ACR Certification of Advanced Proficiency in Cardiac Computed Tomography (COAP)[35], and validation by the Certification Board of Cardiovascular Computed Tomography (CBCCT)[36].

From this derived collection of de-identified patient data, non-consecutive CCTA reports were randomly selected until the predetermined targets of 500 cases (including 200 cases from a prior technology validation report[30]) and 50%-disease prevalence were reached; the disease prevalence in the *Phase 1* study population was deliberately inflated in order to minimize issues related to sample size and model Training. Due to their potential to confound AI algorithm production, this selection process purposely excluded cases with: (1) retained metallic materials related to prior CAD intervention (e.g., stents, bypass graft markers, sternal sutures); or (2) other internal treatment-related materials (i.e., surgical or implanted devices, such as prosthetic valves or pacemakers). However, the image characteristics of the remaining cases were otherwise relatively "real-world" with the inclusion of all image qualities (e.g., equivocal from motion-related artifact) and complicating coronary anatomies (e.g., anomalous origin/course or myocardial bridges).

The final categorization of each selected case as: (1) "Diseased", with non-obstructive (i.e., at most insignificant stenosis < 50% everywhere) or obstructive (stenosis ≥ 50% anywhere) atherosclerotic plaque formation, or as (2) "Normal" (totally free of coronary atherosclerosis), required CCTA examination review/re-review by a Level-3 ACCF/AHA CCTE, ACR-COAP, and CBCCT investigator (RDW or MSG)[34–36].

Ultimately, the desired study population (Age: range 20-91/average 50 years old; Gender: 52% male), consisting of 250 confirmed Diseased cases (125 non-obstructive and 125 obstructive) and 250 confirmed Normal cases, was established [Table 1]. All CCTA examinations had been performed in ED-based (91%) or ambulatory clinic-based (9%) settings; cases of ACS (very uncommon at our institution) were unintentionally unrepresented.

*Coronary Image-Data Curation:*

Image-data curation included ground-truth annotation of the extent of any identified coronary artery atherosclerotic plaque [Appendix B][27] by the most experienced investigator (RDW) who has 35 years of practice in cardiovascular imaging, in addition to the aforementioned credentials.

*Algorithm Development:*

<u>Establishing Classification:</u>

In *Phase 1*, the development of the algorithm for coronary vessel classification was initiated by the establishment of a 3:1:1 distribution[37] of randomly selected study population cases for Training (n = 300), Validation (n = 100), and Testing (n = 100); each of these subsets demonstrated a 1:1 Diseased-to-Normal case ratio [Table 1].

Next, depending on the assigned subset and vessel centerline-extraction condition (i.e., "Plaque-Annotated" *vs* "Atherosclerosis-Free"), corresponding types of Mosaic Projection Views (MPVs)[30] were selectively recruited from cases as follows to establish data partitioning for vessel-based algorithm Training, Validation, and Testing. For the 250 Normal cases, the MPVs of all vessel extractions were included in all three phases of AI algorithm development. On the other hand, for the 250 Diseased cases, while MPVs of all Plaque-Annotated extractions were used, the following were excluded for algorithm development in order to avoid both unrecognized similarities with atherosclerotic vessels and further dataset imbalance favoring Normal[30]: (1) extraction MPVs of atherosclerosis-free branches with "upstream" plaque annotations; and (2) completely atherosclerosis-free extraction MPVs [Table 1].

In order to further promote the relative balance of diseased *vs* non-diseased for algorithm development, a DA strategy was used to randomly permute and augment 6-fold the 394 Plaque-Annotated MPVs in the Training subset[30], thereby creating 2,364 (i.e., 394 x 6) unique representations of atherosclerotic vessels [Table 1]. Traditional DA routines (e.g., random rotation) were also performed[38–41].

On the other hand, DA was not performed on: (1) Atherosclerosis-Free MPVs of Normal cases in the Training subset (n = 2,304); (2) the entire Validation subset of MPVs (both atherosclerotic and non-atherosclerotic); and (3) the entire Testing subset of MPVs (both atherosclerotic and non-atherosclerotic). In the Validation subset, only one MPV was randomly selected per Diseased case (n = 50 MPVs) and per Normal case (n = 50 MPVs), in order to maintain balance and avoid bias in the final algorithm selection[30]. In the vessel extraction-level Testing subset, a single MPV per Plaque-Annotated extraction (n = 125 MPVs), as well as for each Atherosclerosis-Free extraction (n = 1,066 MPVs), was used [Table 1].

Testing on the case level in *Phase 1* (i.e., preliminarily evaluating the algorithm in differentiating between cases with *vs* without coronary atherosclerosis) was performed after full restoration of the Diseased cases. Restoration was accomplished by inclusion of the aforementioned initially excluded MPVs (n = 538 atherosclerosis-free with "upstream" plaque annotations, and n = 353 completely atherosclerosis-free), for completion of Diseased-case image-data sets [Table 1].

<u>Modeling:</u>

A Transfer Learning (TL) strategy[38–41], which has been previously validated[30], was used to help prepare the AI model for CCTA classification [Appendix C].

**Phase 2: Algorithm Trialing and Pre-Deployment Assessment**

*Study Population:*

With local IRB approval, the GUI[27,28] was installed on a standard Picture Archiving and Communication System (PACS) workstation supporting commercially available advanced image-evaluation software (*syngo*.via: Siemens Healthineers, Forchheim, Germany). This allowed Trialing of the developed algorithm in a fashion simulating "real-world" use on a recent standard-of-care series (1/2019-04/2020) of 100 consecutive suitable patients with histories of ED-indicated CCTA evaluation of new-onset chest pain in the setting of low-to-intermediate pretest CAD probability. Out of the total of 106 case records accessed to achieve this goal, only 6 cases were excluded based on: (1) inadequate image quality for AI inference generation although reported as completely free of atherosclerosis by the interpreting physician (n = 4, representing an algorithm workflow-completion level of 100/104 = 96%); (2) CT scanner malfunction resulting in incomplete CCTA examination (n = 1); and (3) presence of a potential AI confounder (n = 1 with implanted defibrillator).

In this *Phase 2* series, each clinical CCTA-examination underwent review/re-review by a Level-3/COAP/CBCCT investigator (RDW or MSG)[34–36] for final categorization as: (1) Diseased, with non-obstructive or obstructive atherosclerotic plaque formation, or (2) Normal. Finally, this 100-consecutive-case study population from the ED (Age: range 22-70/average 45 years old; Gender: 58% male), consisted of 28 confirmed-Diseased and 72 confirmed-Normal cases, reflecting a disease prevalence of 28% atherosclerotic cases (22 non-obstructive and 6 obstructive); there were no confirmed ACS cases in this series.

*Algorithm Trialing:*

During algorithm Trialing in *Phase 2*, the CCTA image-data of each of the 100 study cases was initially reviewed on the aforementioned advanced clinical viewer (i.e., *syngo*.via), as in standard-of-care, for the identification of the cardiac phase with perceived optimal diagnostic quality (often end-diastolic at 65%-75%). Once identified, the following basic steps occurred: (1) the optimal image-data volume was routed for the previously described image processing necessary for algorithm assessment; (2) an algorithm inference was returned (55-80 seconds later, depending on processing and/or network demands) to the GUI with a vessel-by-vessel graphic overlay indicating any detected atherosclerosis on the displayed

### TABLE II
### Statistical results from Testing at vessel-extraction level

| | | *Phase 1* Algorithm Performance Testing: Vessel-Extraction Level (Decision Threshold: 0.5) | | |
|---|---|---|---|---|
| | Formula | Confusion Matrix Values | Calculated Value | 95% CI |
| **Sensitivity** | $\frac{TP}{TP+FN}$ | 107 / (107+18) | 85.60% | 78.20% to 91.24% |
| **Specificity** | $\frac{TN}{FP+TN}$ | 988 / (78+988) | ***92.68%*** | 90.95% to 94.17% |
| **PPV** | $\frac{TP}{TP+FP}$ | 107 / (107+78) | 57.84% | 52.27% to 63.22% |
| **NPV** | $\frac{TN}{FN+TN}$ | 988 / (18+988) | ***98.21%*** | 97.28% to 98.83% |
| **Accuracy** | $\frac{TP+TN}{TP+FP+FN+TN}$ | (107+988) / (107+78+18+988) | 91.94% | 90.25% to 93.42% |

(FN=False-Negative, FP=False-Positive, NPV=Negative Predictive Value, PPV=Positive Predictive Value, TN=True-Negative, and TP=True-Positive)

### TABLE III
### Statistical results on preliminary Testing of randomly selected cases

| | | *Phase 1* Algorithm Performance Testing: Randomly Selected-Case Level | | |
|---|---|---|---|---|
| | Formula | Confusion Matrix Values | Calculated Value | 95% CI |
| **Sensitivity** | $\frac{TP}{TP+FN}$ | 49 / (49+1) | 98.00% | 89.35% to 99.95% |
| **Specificity** | $\frac{TN}{FP+TN}$ | 20 / (30+20) | ***40.00%*** | 26.41% to 54.82% |
| **Disease Prevalence** | $\frac{TP+FN}{TP+FP+FN+TN}$ | (49+1) / (49+30+1+20) | 50.00% | - |
| **PPV** | $\frac{TP}{TP+FP}$ | 49 / (49+30) | 62.03% | 56.48% to 67.27% |
| **NPV** | $\frac{TN}{FN+TN}$ | 20 / (1+20) | ***95.24%*** | 73.61% to 99.31% |
| **Accuracy** | $\frac{TP+TN}{TP+FP+FN+TN}$ | (49+20) / (49+30+1+20) | 69.00% | 58.97% to 77.87% |

(FN=False-Negative, FP=False-Positive, NPV=Negative Predictive Value, PPV=Positive Predictive Value, TN=True-Negative, and TP=True-Positive)

### TABLE IV
### Statistical results from Trialing on consecutive cases

| | | *Phase 2* Algorithm Performance Trialing: Consecutive-Case Level | | |
|---|---|---|---|---|
| | Formula | Confusion Matrix Values | Calculated Value | 95% CI |
| **Sensitivity** | $\frac{TP}{TP+FN}$ | 26 / (26+2) | 92.86% | 76.50% to 99.12% |
| **Specificity** | $\frac{TN}{FP+TN}$ | 36 / (36+36) | ***50.00%*** | 37.98% to 62.02% |
| **Disease Prevalence** | $\frac{TP+FN}{TP+FP+FN+TN}$ | (26+2) / (26+36+2+36) | 28.00% | - |
| **PPV** | $\frac{TP}{TP+FP}$ | 26 / (26+36) | 41.94% | 35.93% to 48.19% |
| **NPV** | $\frac{TN}{FN+TN}$ | 36 / (2+36) | ***94.74%*** | 82.27% to 98.59% |
| **Accuracy** | $\frac{TP+TN}{TP+FP+FN+TN}$ | (26+36) / (26+36+2+36) | 62.00% | 51.75% to 71.52% |

(FN=False-Negative, FP=False-Positive, NPV=Negative Predictive Value, PPV=Positive Predictive Value, TN=True-Negative, and TP=True-Positive) coronary artery tree (composite 3D display of centerline-extracted coronary vessels); and (3) interpreter adjudication (accept/reject) with potential CCTA revisiting on the GUI, as justified on indicated likelihood of coronary atherosclerosis[42] [Appendix D].

### D. Statistical Analysis

In *Phase 1*, Testing of the performance of the AI algorithm for vessel-extraction classification (Plaque-Annotated *vs* Atherosclerosis-Free) in the 500-random case study population was evaluated using the Area Under the Receiver

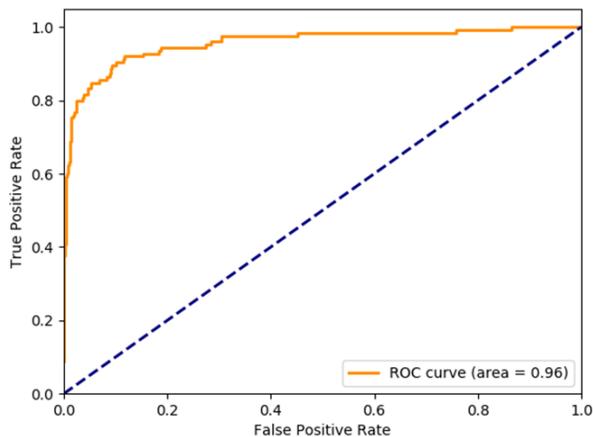

Fig. 1. Area Under the Receiver Operating Characteristic Curve (AUC-ROC) curve from Testing of vessel-based algorithm performance.

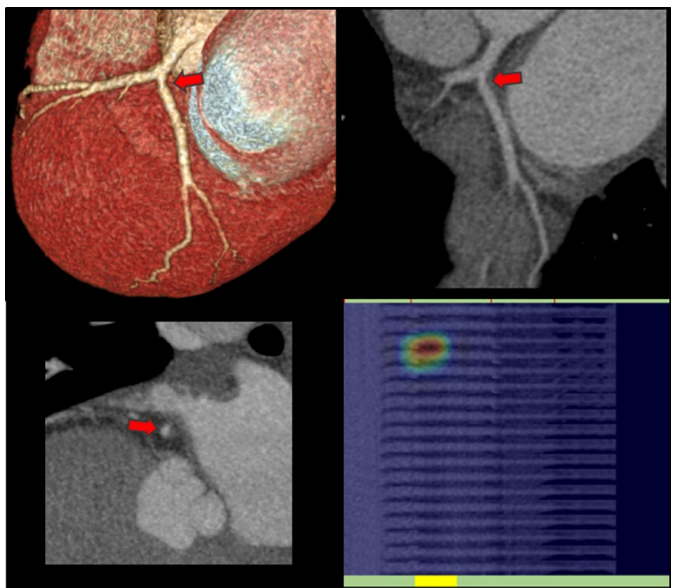

Fig. 2. Single false-negative case-level determination during preliminary algorithm Testing

A focal mild (24-49%) atherosclerotic stenosis [Arrows] immediately after the origin of the Left Circumflex coronary artery, with bifurcation of the Left Main Coronary Artery, is shown on the volume-rendered (Upper Left) image and multi-planar reconstructions (Upper Right and Lower Left). A VGG16-based saliency map (Lower Right) superimposed on the MPVs (unique 18 x 1 matrices of arranged 2D projections–newer version) of this case demonstrates a focal "hot spot" region indicating a discriminative area by the algorithm (corresponding to manually demarcated segment of mild atherosclerosis [yellow]) along projection images from the vessel extraction. However, this "hot spot" went under-represented due to the predicted probability for plaque of only 0.02, below the diagnostic threshold used (i.e., 0.5).

Operating Characteristic Curve (AUC-ROC) methodology[43]. For standard analysis of Testing results (especially NPV), confusion matrices were used to reflect expert "ground-truth" determinations vs AI-algorithm predictive determinations of atherosclerosis presence vs absence at both the vessel-extraction level and case level[38–41]. As previously mentioned, case-level Testing was performed after Diseased case restoration by addition of previously excluded: (1) extraction MPVs of atherosclerosis-free branches with "upstream" plaque annotations; and (2) completely atherosclerosis-free extraction MPVs.

In *Phase 2*, algorithm Trialing was performed on the 100-consecutive-case study population from the ED, using confusion matrices to reflect expert determinations *vs* algorithm determinations of atherosclerosis presence *vs* absence on a per-case basis.

## III. RESULTS

### Phase 1: Algorithm Development and Preliminary Testing

In the *Phase 1* 500-case study population, expert ground-truth was established for 3,420 Atherosclerosis-Free vessel-centerline extractions from the 250 Normal cases, as well as for 569 (i.e., 394 + 50 + 125) Plaque-Annotated extractions from the 250 Diseased cases [Table 1]. While all 3,989 (i.e., 3,420 + 569) vessel extractions were converted to MPVs for Training, Validation, and Testing during algorithm development; only the 394 Plaque-Annotated MPVs in the Training subset underwent DA (creating 2,364 different representations of atherosclerotic arteries/branches).

The optimized algorithm developed in *Phase 1* revealed potentially very high performance at the vessel-extraction level, with an AUC-ROC of 0.96 measured [Figure 1]. When the Testing results were reviewed further [Table 2], the confusion matrix statistics confirmed the potential strength of the algorithm for total exclusion of atherosclerosis per vessel extraction, with calculated high NPV of 98.21% (95%CI 97.28% - 98.83%) and Specificity of 92.68% (95%CI 90.95% - 94.17%).

Nevertheless, with the ultimate goal being to assess AI in screening for the absence of atherosclerosis anywhere within a chest-pain patient's coronary artery system on CCTA, case-level evaluation was the major focus. To that end, the developed algorithm was applied in *Phase 1* to per-case evaluation in the Testing subset (including 50 restored Diseased cases and 50 Normal cases) [Table 3]. The results revealed a case-level NPV of 95.24% (95%CI 73.61% to 99.31%) and Specificity of 40.00% (95%CI 26.41% to 54.82%). There was False-Negative (FN) determination in only 1 case which demonstrated a single atherosclerotic narrowing of mild severity (25-49% stenosis) immediately after the origin of the vessel [Figure 2].

### Phase 2: Algorithm Trialing and Pre-Deployment Assessment

When the developed algorithm was applied to per-case Trialing in the *Phase 2* study population (including 28 Diseased cases and 72 Normal cases), the workflow experience was operationally successful [Figure 3] and the results resembled *Phase 1* preliminary case-level results [Table 3]. NPV of 94.74% (95%CI 82.27% to 98.59%) and Specificity of 50.00% (95%CI 37.98% to 62.02%) were calculated at the case level in *Phase 2*. The results revealed FN determination in only two cases; both demonstrated single-vessel (both Left Anterior Descending coronary artery) non-obstructive atherosclerosis (<25% and 25-49% stenosis) immediately beyond branch origins [Figure 4]. Although not leading to case-level FN determinations, there were 5 additional vessel extraction-level FN determinations (3 Right

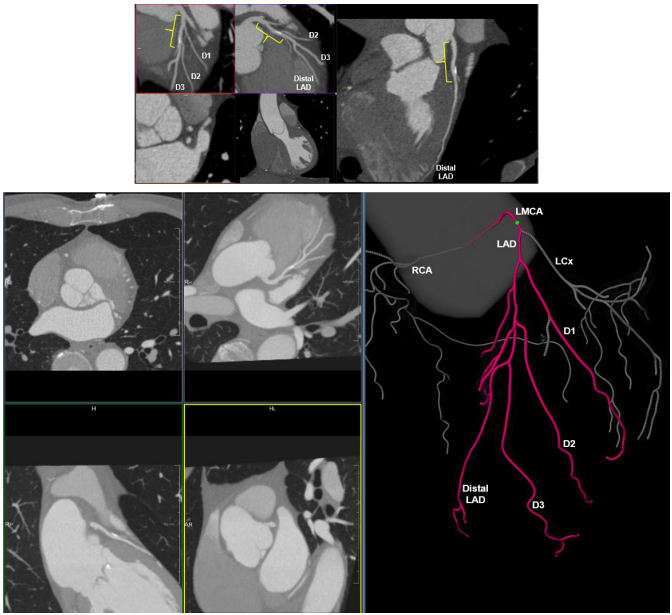

Fig. 3. Algorithm inference feedback on detected atherosclerosis.

Non-obstructive atherosclerosis [Yellow Brackets] spanning the levels of the Left Main Coronary Artery [LMCA] to the proximal-mid Left Anterior Descending coronary artery [LAD] is demonstrated on multi-planar-reconstruction and maximum-intensity-projection images from a commercial clinical viewer (Above). The GUI display (Below) supports comparable interactive-viewing capabilities, as well as a graphic overlay of inference results from plaque detection [Red] on the vessel-centerline extracted coronary tree [Gray]. Algorithm-based detection of atherosclerosis in the LMCA, Proximal LAD (from LMCA bifurcation to origin of first Diagonal branch [D1]) and mid LAD (from D1 original to origin of second Diagonal branch [D2]) conveys distally to the "down-stream" non-diseased distal LAD and D1-D3. Without disease detected in the Left Circumflex coronary artery [LCx] or Right Coronary Artery [RCA], no overlays are applied to their vessel distributions.

Coronary Artery, including 2 main portion and 1 posterior descending branch; and 2 Left Circumflex coronary artery, including 1 main portion and 1 obtuse marginal).

## IV. Discussion

Building on a shared view that *non-obstructive coronary artery atherosclerosis detected during CCTA evaluation of chest-pain should not be ignored*[44,45], this work was focused on the added diagnostic value of AI in assisting the interpreting physician. The importance of *totally excluding coronary atherosclerosis aided by AI*, particularly in typical ED chest-pain patients of low-to-intermediate pretest CAD probability, has implications beyond facilitating discharging to home, operational efficiencies, and financial benefits; failure to detect even mild coronary plaque formation can leave the patient in jeopardy of a future MACE[12,13]. Therefore, the confident confirmation of a completely normal-appearing CCTA to help determine minimal risk in an ED chest-pain patient should be considered to be an important objective along with the detection of CAD.

Based on the results of this study, AI could soon play a significant role in assisting the interpreting physician in achieving this objective of coronary atherosclerosis exclusion.

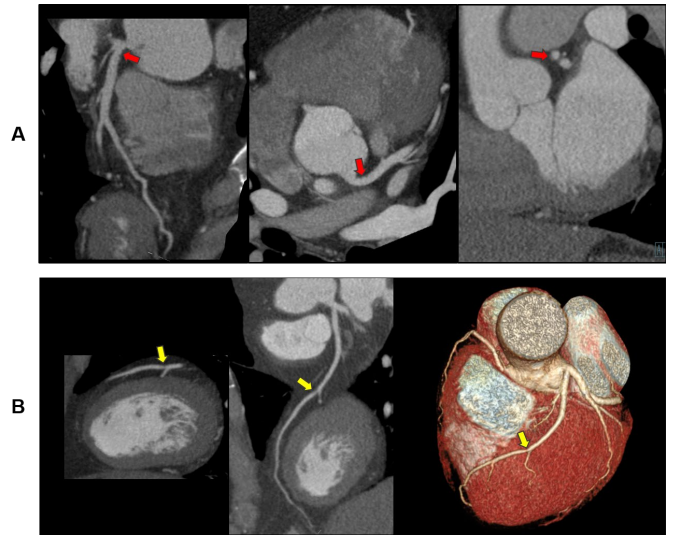

Fig 4. False-negative case-level determinations during algorithm Trialing

False-negative determinations were observed in only 2 cases; both demonstrated single-vessel non-obstructive atherosclerosis of the Left Anterior Descending coronary artery (LAD) immediately beyond a branching point. In case A (Above), curved and straight multi-planar reconstructions show minimal (< 25%) atherosclerotic stenosis [Red Arrows] in the LAD after its origin from the Left Main Coronary Artery; an adjacent very small Ramus Intermedius branch is noted. In case B (Below), straight and curved multi-planar, as well as Volume Rendered (Right), reconstructions show focal atherosclerosis causing mild (25-49%) stenosis [Yellow Arrows] of the mid LAD just beyond the origin of the first Diagonal branch.

First, they indicate a very high performance of the developed AI algorithm at the vessel level (i.e., AUC-ROC = 0.96); the small number of FN determinations resulted in very high NPV (i.e., 98%) and high Specificity (i.e., 93%) per vessel extraction. Multiple prior validation studies have also demonstrated a strong ability of CCTA to exclude atherosclerosis per vessel[4,5,46,47], but this has necessitated labor-intensive manual analysis which may be difficult to support, for the aforementioned reasons, within an ED environment.

More pertinent to the practical goal of *totally excluding coronary atherosclerosis per chest-pain patient* for their clinical disposition and treatment planning in the ED, our AI algorithm demonstrated very high NPV (95%) in both the *Phase 1* (disease prevalence of 50%) and *Phase 2* (disease prevalence of 28%) study populations, thereby indicating that the post-test probability of an ED chest-pain patient having no coronary atherosclerosis, given a *completely atherosclerosis-free* AI-evaluated CCTA result, approaches absolute, with low likelihood of a FN determination[48]. On the other hand, our previously described high Specificity per-vessel extraction was not observed in either case population (40% and 50%, respectively), reflecting the equivocal ability of our algorithm to alone correctly designate an ED chest-pain patient without coronary atherosclerosis as being *completely* atherosclerosis-free, with significant potential for False Positive determinations warranting thorough manual evaluations for suspected plaque formations[48]. Even when CCTA examinations are manually interpreted, decline in Specificity per-patient, compared to per-segment or per-vessel basis, has been reported[4,5,46,47].

Hence, these results do not endorse the use of our algorithm for autonomous AI-based CCTA interpretation; this was not anticipated, especially when it is recognized that expert review is still essential due to potential for other symptom-producing abnormalities, such as congenital coronary artery anomalies. However, they do help validate its use as a pre-view or post-view assistant to the interpreting physician in preventing mistakes of diagnostic omission that might otherwise lead to unrecognized risks of MACE by failing to detect coronary atherosclerosis prior to ED discharge without plans for further evaluation or treatment. In addition, to that end, the results indicate that our algorithm workflow is robust with: (1) reliable performance when image quality is not poor (100 of 104 = 96% of real-world cases without obvious confounders demonstrated algorithm workflow completion with inference feedback); (2) acceptable turnaround time (55-80 seconds) of graphic inference results; and (3) post-inference opportunity for the interpreting physician to adjudicate inference results by easy interactive re-review of CCTA images within the same clinically capable GUI.

**Limitations**

Limitations in Study Population:

Unlike prior reports on the use of AI for anatomic/physiologic detection or characterization of known coronary artery atherosclerotic plaque[20–23], this project focused on the AI-based exclusion any atherosclerosis. Apparent deficiencies in autonomous detection of atherosclerosis (per-case positive predictive values of 42-62%) are likely related to both *Phase 1* and *Phase 2* study populations having included: 1. relatively "real-world" examinations spanning all image qualities (rendering workflow incomplete in 4 *Phase 2* cases with very poor imaging); and 2. a high prevalence of vessel extractions which were atherosclerosis-free or had non-obstructive (often very mild) atherosclerosis, partially ameliorated in *Phase 1* but left in "real world" status in *Phase 2*. In order to correct for this latter issue for algorithm development in *Phase 1*, disease prevalence was purposefully inflated to 50% during case selection in order to minimize the sample-size concern; consequently, the calculated sample-size need (i.e., 146 cases) was reduced to a level surpassed by the combined Training-Validation subsets (i.e., 400 cases) and approximated by the Testing subset (i.e., separate 100 cases). Data balancing was also employed to ensure disease representation throughout the vessel-based algorithm development process[30].

Limitations in Technology:

A variety of technique options are available for vessel-centerline extraction, and each has relative inherent strengths and deficiencies[49], beyond practical restrictions related to poor image quality, reduced intravascular enhancement, and severe vessel tortuosity. Technical limitations of the fully automatic centerline methodology applied in this study have been described[27,30], and cumulatively these limitations may have adversely affected vessel extraction and ultimately algorithm development.

In addition, known shortcomings with centerline recognition at vessel bifurcations may have contributed to the occurrence of missed AI-algorithm detection of focal atherosclerotic plaques in this study[50]. Nevertheless, the three depicted instances leading to FN cases involved failed detection of non-obstructive degrees of atherosclerosis, regarded to be "negative" in many prior studies of CCTA in ED chest pain-patients[5]. Also, planned work to support the continuous learning process of our AI algorithm (in contrast to traditional computer aided diagnosis[38–41]) is expected to improve its future performance, thereby reducing the potential for such FN inference results.

Lastly, there are impending clinical implementation issues related to the current algorithm workflow, including its: (1) requirement for manual selection of the cardiac phase with optimal image quality which may eventually be overcome with technical options (e.g., automatic phase selection[51], optimization via motion correction[52]); 2. inability to graphically display local regions of highest probability for plaque detection, or provide for probability threshold adjustment, which may be overcome with 3D and/or neural network-dependent approaches in the future; and (3) ongoing dependencies on interfacing of commercially based (e.g., centerline technology) and locally developed approaches (e.g., MPV production), also potentially benefitting from the same approaches.

## V. Conclusion

This report describes the methodical development of an AI algorithm, and supporting workflow, which appears to have reached a level of performance conducive to clinical deployment for assisting an interpreting physician in confirming the *total absence of any coronary artery atherosclerosis* on CCTA during a chest-pain patient evaluation in the ED setting. The AI application's validated combination of: (1) very high NPV at the case level; (2) efficient return of results from algorithm inference; and (3) graphic overlay of inference results on the coronary artery tree display, with allowance for subsequent interactive review of CCTA images within the same "user friendly" GUI, promotes its readiness for clinical implementation. It also anticipates the expectation for support of continuous algorithm improvement, inherent in AI. Provision of such AI-based assistance to the interpreting physician in confidently *discounting coronary atherosclerosis as the cause of chest pain* in an ED chest-pain patient could help to avoid under-diagnosis or under-treatment due to human error.

49. Schaap M, Metz CT, van Walsum T, et al. Standardized evaluation methodology and reference database for evaluating coronary artery centerline extraction algorithms. *Med Image Anal*. 2009;13(5):701-714. doi:10.1016/j.media.2009.06.003

50. Wang Y, Liatsis P. 3-D quantitative vascular shape analysis for arterial bifurcations via dynamic tube fitting. *IEEE Trans Biomed Eng*. 2012;59(7):1850-1860. doi:10.1109/TBME.2011.2179654

51. Hadjiiski L, Liu J, Chan HP, et al. Best-Quality Vessel Identification Using Vessel Quality Measure in Multiple-Phase Coronary CT Angiography. *Comput Math Methods Med*. 2016;2016. doi:10.1155/2016/1835297

52. Balaney B, Vembar M, Grass M, et al. Improved visualization of the coronary arteries using motion correction during vasodilator stress CT myocardial perfusion imaging. *Eur J Radiol*. 2019;114:1-5. doi:10.1016/j.ejrad.2019.02.010

# APPENDICES

## Appendix A

Because all vessel-centerline extractions originated at the transition from the aortic root to either the Left Main Coronary Artery or Right Coronary Artery (standard in commercial coronary artery image-processing systems), the sharing of an "up-stream" arterial segment between "down-stream" segments of the same primary artery or its branches is common. In this study, while the identification of each vessel extraction was based on the vessel of termination, its classification as "Plaque-Annotated" or "Atherosclerosis-Free" was established by the presence or absence, respectively, of atherosclerosis anywhere along the vessel extraction, even when atherosclerosis was located only before the true origin of a branch [Figure A].

## Appendix B

The GUI-based manual annotation process involved regional color-coding of basic stenosis grade (i.e., non-obstructive vs obstructive) along the vessel courses as shown on rotatable 3D branching arterial trees, representing the combined display of the multiple vessel-centerline extractions per case[27] [Figure B1].

The annotated vessel extractions (i.e., Plaque-Annotated) were then converted to 2D Mosaic Projection View (MPV) displays in order to facilitate Data Augmentation (DA) of the size of the atherosclerotic-vessel component of Training for algorithm development[30]. DA is accomplished by re-ordering projections, thereby creating new MPVs per vessel used to augment the Training subset [Figure B2]. Through its improvements to both modeling and diseased *vs* non-diseased balance within the Training subset[37], this DA methodology

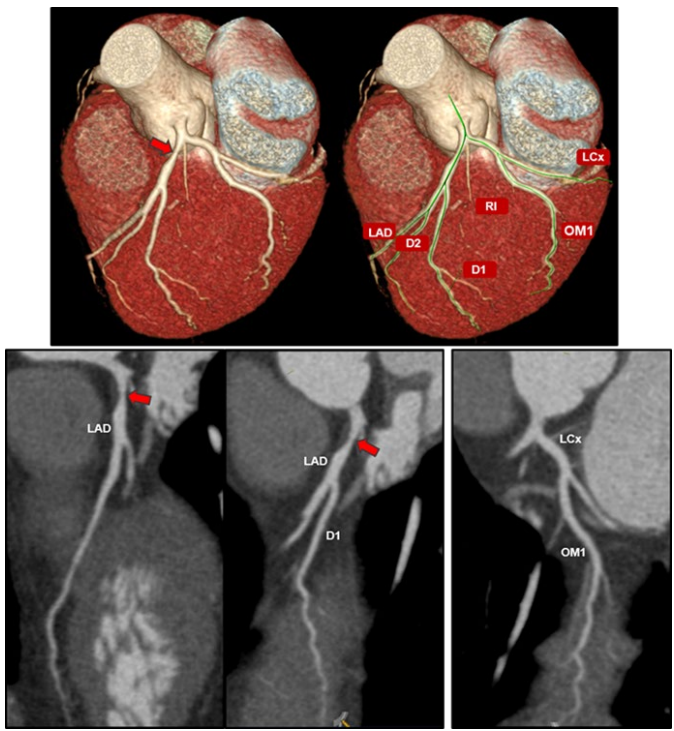

Fig A. Classification of vessel-centerline extractions

Significant luminal narrowing [Arrows] of the Left Anterior Descending coronary artery [LAD] is demonstrated on both a volume-rendered image (Above, with automatically extracted centerlines indicated on Right) and curved multi-planar images along the vessel centerlines (Below). The "up-stream" atherosclerotic proximal LAD dictates the classification of both the "down-stream" atherosclerosis-free mid-distal LAD and first Diagonal branch [D1] as "Plaque-Annotated". Hence, D1 is considered a diseased vessel course due to the proximal LAD atherosclerosis; for it to be considered Atherosclerosis-Free requires all segments proximal to the branch to also be free of atherosclerosis. On the other hand, the Left Circumflex coronary artery [LCx] and its first Obtuse Marginal branch [OM1] are entirely Atherosclerosis-Free along their vessel extractions. [D2 = second Diagonal branch; RI = Ramus Intermedius branch]

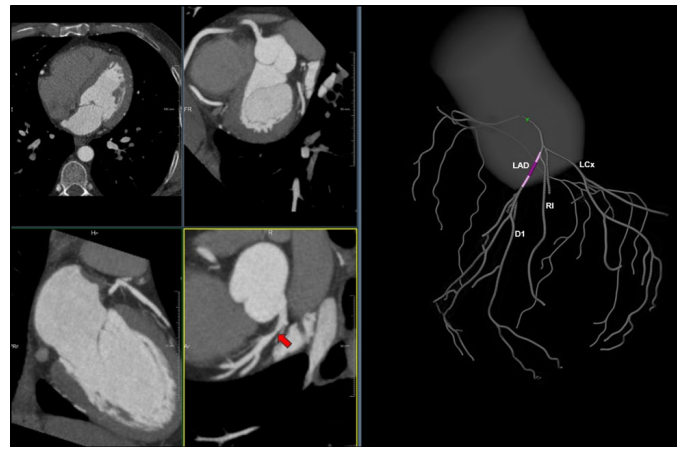

Fig B1. GUI for ground-truth annotation of coronary arteries and branches

The same case as represented in Figure A helps to demonstrate the subsequent GUI-based image-data curation with ground-truth annotation of coronary artery atherosclerotic plaque extent. This process involves regional color-coding of basic stenosis grade of the previously described significant stenosis [Arrow] of the proximal Left Anterior Descending coronary artery [LAD]; both non-obstructive (< 50% luminal narrowing [Light Pink]) and obstructive (≥ 50% luminal narrowing [Dark Pink]) portions of the plaque are shown as segmental coatings on the 3D branching arterial tree [D1 = first Diagonal branch; LCx = Left Circumflex coronary artery; RI = Ramus Intermedius branch]

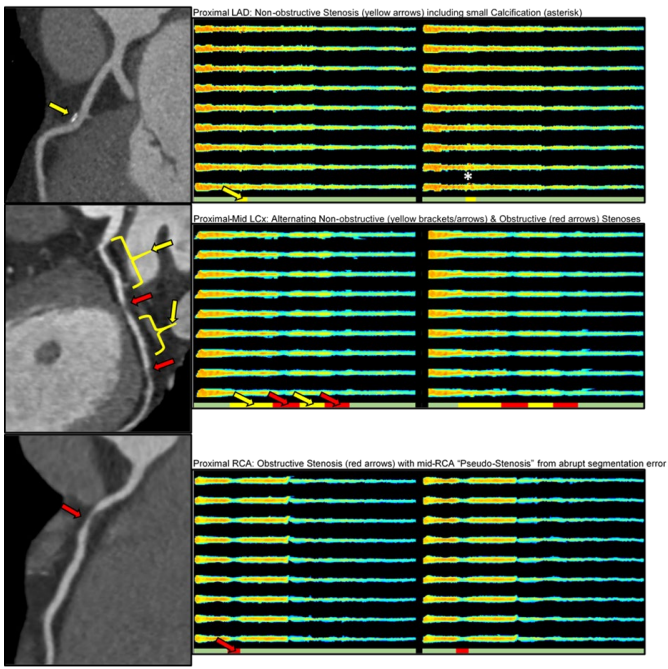

Fig B2. Mosaic Projection View (MPV) representations of disease-annotated vessel extractions

Three MPVs (unique 9 x 2 matrices of arranged 2D projections – original version [30]) of coronary vessel atherosclerosis with variable stenosis are displayed as color maps (Right); corresponding plaques are shown on curved CCTA reconstructions (Left). [LAD = Left Anterior Descending, LCx = Left Circumflex, RCA = Right Coronary Artery].

enhanced AI algorithm development, especially when combined with Transfer Learning (TL) using pre-trained weighting[30,37–41].

### Appendix C

Inception-V3 [https://cloud.google.com/tpu/docs/inception-v3-advanced] served as the base convolutional neural network. Model weights pre-trained on ImageNet [http://www.image-net.org/] were used for TL during algorithm Training[30,38–41]. To refine modeling for CCTA datasets, the final Inception-V3 layer was replaced by a fully connected 1024-node layer with a rectified linear unit (25% dropout to avoid overfitting), followed by sigmoid output function for binary classification[30] [Figure C].

Training utilized the Keras library [https://keras.io/] with a TensorFlow-1.8 backend [https://www.tensorflow.org/]. Initial learning rate was 0.001 on a stochastic gradient descent optimizer (decay factor 1e-6, with momentum 0.900 and mini-batch size 8); re-training was terminated at 120 epochs. A binary cross-entropy loss function was monitored during Training, and the resulting model was saved only if there was improvement in the Validation accuracy[30].

### Appendix D

For *Phase 2*, Clara software stack [https://developer.nvidia.com/clara-medical-imaging] was used because of its support of: (1) an interface to Digital Imaging and Communications in Medicine (DICOM) standards [www.dicomstandard.org]; (2) a durable deployment of Kubernetes [https://kubernetes.io/docs/tasks/manage-gpus/scheduling-gpus/#clusters-containing-different-types-of-nvidia-gpus]; and (3) seamless modular integration with existing image-visualization/analysis tools, including a PACS or a viewer.

The integrated workflow for pre-deployment of the algorithm developed in *Phase 1* included the following steps: (1) the interpreting physician opened the desired CCTA image-data in a commercial clinical viewer [https://www.siemens-healthineers.com/en-us/medical-imaging-it/advanced-visualization-solutions/syngovia] in order to select the most diagnostically optimal cardiac phase to undergo evaluation by the algorithm; (2) the selected volume was manually forwarded to the DICOM server from where it was sent to the server supporting the GUI [https://www.mevislab.de/] for coronary artery image-processing (including vessel-centerline extraction followed by production of straightened Multi-Planar Reformations (MPRs); (3) following conversion, straightened-MPRs were sent to a Clara Deploy DICOM Adapter [https://docs.nvidia.com/clara/deploy/ngc/DicomAdapter.html] which monitors incoming DICOM images and initiates classification by the algorithm; (4) Clara Deploy platform, hosting a TensorRT Inference server [https://docs.nvidia.com/deeplearning/sdk/triton-inference-server-guide/docs/index.html], is prompted to make predictions; (6) inference results are updated in the database

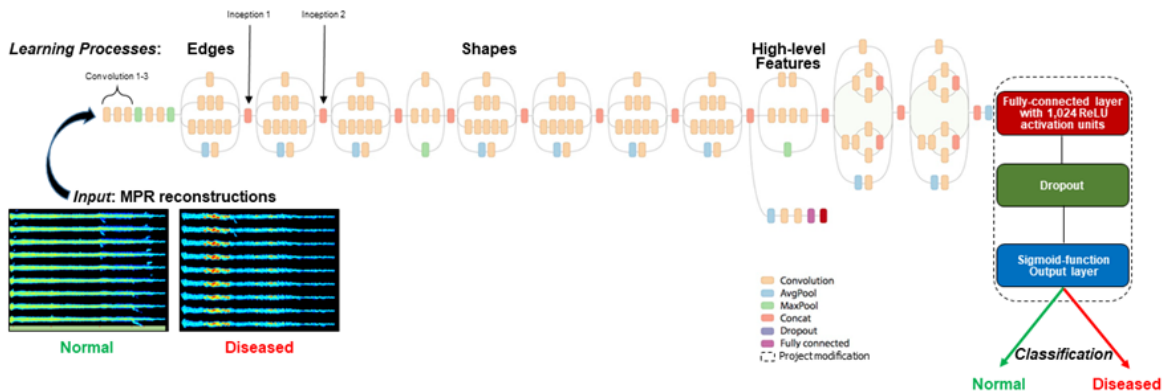

Fig C. Modified Inception-V3 convolutional neural network used for algorithm development

[https://www.mongodb.com/], signaling readiness for feedback to the interpreting physician; and (7) inference probability values exceeding the threshold (0.5) are graphically displayed as overlays on the coronary artery tree displayed by the GUI.

Using a continuous model integration strategy, an additional model can be evaluated by updating the CLARA Deploy server, via the backend server without workflow disruption, for future selection by the interpreting physician. The application described in *Phase 2* and pre-deployed for simulated clinical use is illustrated below [Figure D].

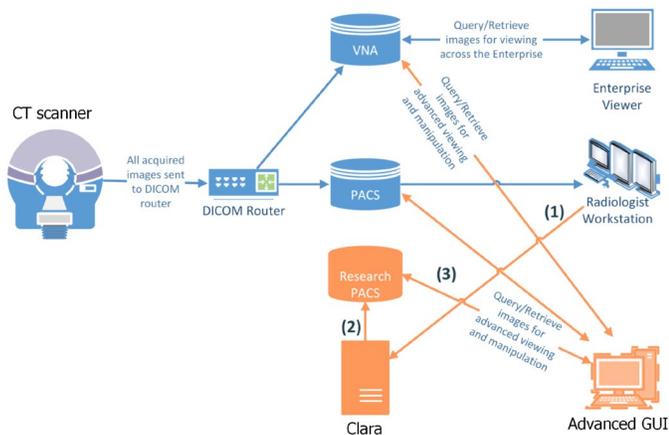

Fig D. Algorithm Pre-Deployment Architecture

While blue components reflect a traditional Radiology workflow, orange components are enhancements for AI algorithm deployment with graphic inference feedback to the interpreting physician via the GUI